\documentclass[10pt,conference,letterpaper]{IEEEtran}
\IEEEoverridecommandlockouts
\pdfoutput=1
\usepackage[T1]{fontenc}
\usepackage{cite}
\usepackage{amsmath,amssymb,amsfonts}
\usepackage{algorithm}
\usepackage[noend]{algpseudocode}
\usepackage{mathtools}
\usepackage{graphicx}
\usepackage{textcomp}
\usepackage{dcolumn} 
\usepackage{bm} 
\usepackage{braket}
\usepackage{hyperref} 
\hypersetup{
    colorlinks = true,
    citecolor = magenta,
    linkcolor = purple
}
\def\BibTeX{{\rm B\kern-.05em{\sc i\kern-.025em b}\kern-.08em
    T\kern-.1667em\lower.7ex\hbox{E}\kern-.125emX}}

\newcommand{\thrgs}{\tau_{\text{half-RGS}}}
\usepackage[svgnames, table]{xcolor}
\usepackage{tabularx, makecell, linegoal}

\usepackage{ifthen}

\newboolean{ShowComments}
\setboolean{ShowComments}{true}  
\ifthenelse{\boolean{ShowComments}}%
	{
		\newcommand{\ColorComment}[3]{%
				{\colorbox{#1}{\color{White}   \textsf{\textbf{#2}}} \textcolor{#1}{#3}}}

	}%
	{
		\newcommand{\ColorComment}[3]{}

	}%

\definecolor{rdvcolor}{rgb}{0,0.5,0}
\definecolor{satohcolor}{RGB}{254,0,0}
\definecolor{michalcolor}{RGB}{255,127,80}
\definecolor{naphanncolor}{RGB}{112, 51, 173}

\begin{document}
\bstctlcite{IEEEexample:BSTcontrol}

\title{
Integrating Entanglement Purification into All-Photonic Quantum Repeaters
\thanks{
This work was supported by the JST [Moonshot R\&D program] under Grant Number [JPMJMS226C].
}
}

\author{
\IEEEauthorblockN{
Naphan Benchasattabuse\IEEEauthorrefmark{1}\IEEEauthorrefmark{3},
Michal Hajdu\v{s}ek\IEEEauthorrefmark{1}\IEEEauthorrefmark{3},
and Rodney Van Meter\IEEEauthorrefmark{2}\IEEEauthorrefmark{3}}\\

\IEEEauthorblockA{\IEEEauthorrefmark{1}\textit{Graduate School of Media and Governance, Keio University Shonan Fujisawa Campus, Kanagawa, Japan}}
\IEEEauthorblockA{\IEEEauthorrefmark{2}\textit{Faculty of Environment and Information Studies, Keio University Shonan Fujisawa Campus, Kanagawa, Japan}}
\IEEEauthorblockA{\IEEEauthorrefmark{3}\textit{Quantum Computing Center, Keio University, Kanagawa, Japan}\\
\{whit3z,michal,rdv\}@sfc.wide.ad.jp}
}

\thispagestyle{plain}
\pagestyle{plain}
\maketitle

\begin{abstract}
We propose a purification-enhanced all-photonic quantum repeater scheme based on repeater graph states (RGS) framework that leverages the recently proposed half-RGS building block.
This framework addresses a longstanding open question--how to naturally integrate entanglement purification with an all-photonic scheme--by enabling long-distance purification without disrupting the core design.
Our framework utilizes optimistic purification performed directly on the half-RGS primitives across long distances without waiting for heralding outcomes.
The overhead is modest: the RGS generation slows down proportionally with the number of purification rounds, and each round requires only one additional quantum emitter per half-RGS source.
However, since the generation time is negligible compared to the end-to-end communication delay, the total latency remains effectively dominated by communication time, similar to frameworks without purification.
Our framework enables flexible purification scheduling along the connection path, making it compatible with memory-based strategies, a rich body of research on purification scheduling and optimization that was previously thought inapplicable to the RGS scheme.
Through numerical evaluation, we compare the performance of our framework with purification between memories at end nodes.
\end{abstract}


\section{Introduction}
\label{sec:introduction}

The realization of quantum networks~\cite{rdv-quantum-networking-book} and ultimately the Quantum Internet~\cite{rfc9340, wehner-vision-road-ahead, rdv-qi-architecture, michal-rdv-qc-book} relies on the efficient distribution of entanglement across distant nodes.
Due to the inherent limitations of direct transmission over long distances~\cite{plob-bound, tgw-bound}--primarily photon loss and decoherence--quantum repeaters~\cite{dur-repeater, azuma-rmp-repeater-review} play a crucial role in enabling scalable quantum communication.
Traditional quantum repeaters employ quantum memories to store entangled states, allowing for entanglement swapping~\cite{zukowski-entanglement-swap}--the process of extending entanglement over longer distances by splicing shorter-length entangled pairs--and entanglement purification (also known as distillation), which enhances fidelity by consuming multiple noisy entangled pairs to distill a smaller number of higher-quality ones~\cite{bennett-purification, deutsch-purification, briegel-repeater, dur-briegel-purification-review}.
However, the requirement for long-coherence time quantum memories remains a significant technological challenge.

An alternative approach is the all-photonic quantum repeater~\cite{azuma-rgs}, which eliminates the need for quantum memories at intermediate nodes by leveraging highly entangled photonic graph states, specifically repeater graph states (RGS).
In this paper, we focus on the all-photonic repeater based on the RGS~\cite{azuma-rgs}, referred to as the RGS scheme.
All-photonic repeaters offer high entanglement generation rates and inherent robustness against photon loss due to built-in redundancy~\cite{varnava-counterfactual-measurement}, making them attractive for scalable architectures that minimize reliance on quantum memories.
However, quantum memories are still required at the end nodes to store the distributed entangled states~\cite{naphan-rgs-protocol}, as Pauli frame correction information--necessary to ensure the Bell pairs are in the correct state--is only available after the distribution process.

Despite these advantages, all-photonic repeaters are ultimately limited by the fidelity of physical qubits, as not all parts of the RGS are error-tolerant.
While their strong tolerance to loss enables near-deterministic distribution, accumulated errors over long distances still degrade the final fidelity of distributed Bell pairs.
High-fidelity entanglement is essential for many quantum networking applications, which require strict fidelity thresholds, including quantum key distribution~\cite{bb84-conf, xu-pan-qkd-review}, private delegated computation~\cite{bfk-blind-protocol, fitzsimons-verifiable-blind-qc}, distributed quantum computing~\cite{cleve-buhrman-entanglement-for-communication, cirac-distributed-qc, gottesman-chuang-universality-teleportation, barral-review-distributed-qc}, and quantum-enhanced metrology~\cite{degen-sensing, ge-distributed-metrology, giovannetti-advances-in-metrology}.
As a result, entanglement purification becomes necessary to enhance the fidelity of the generated Bell pairs.

A natural approach to entanglement purification in all-photonic networks is to perform two-way purification~\cite{bennett-purification, deutsch-purification} at the end nodes after establishing long-distance entangled states.
While straightforward, this strategy introduces latency due to classical communication, and memory decoherence during this time can reduce both the fidelity and success rates of purification--ultimately lowering the end-to-end distribution rate.
Another possibility is measurement-based purification~\cite{zwerger-measurement-based-repeater}, which uses purification-embedded graph states to purify link-level entanglements.
However, this method is restricted to adjacent nodes, and its success probability is much lower than purification on memories.
Consequently, multiple purification rounds significantly reduce the generation rate.
Thus, integrating efficient and flexible purification schedules into the RGS scheme without compromising its attractive features remains a vital challenge~\cite{azuma-rgs}.

\begin{figure*}[!ht]
    \centering
    \includegraphics[width=\textwidth]{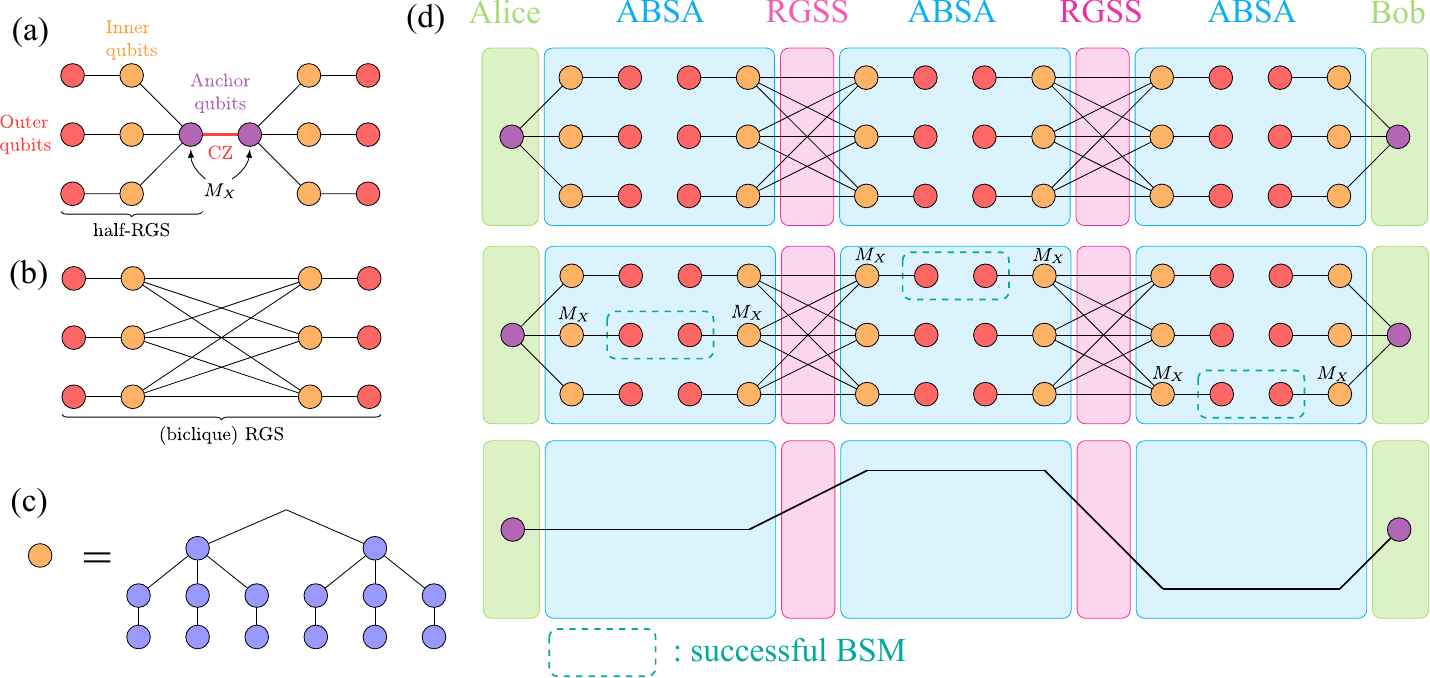}
    \caption{
        An overview of the RGS scheme, illustrating the node types and qubits involved.
        (a) A half-RGS with three arms, where purple, orange, and red circles represent anchor, inner, and outer qubits, respectively.
        Two half-RGSs can be combined into an RGS by applying a CZ gate and performing X-basis measurements on the anchor qubits.
        (b) A complete RGS formed by joining two half-RGSs at their anchor qubits.
        (c) An example of a tree-encoded inner qubit with branching parameters (2,3,1).
        (d) A schematic of Bell pair distribution in the RGS-based repeater chain.
        End nodes (Alice and Bob) and all RGS sources (RGSS) along the path generate and transmit their half-RGSs to adjacent advanced Bell state analyzer (ABSA) nodes.
        For clarity, the conversion of half-RGSs into full RGSs at RGSS is omitted.
        Bell state measurements (BSMs) are performed on pairs of outer qubits arriving from opposite sides of each ABSA.
        Each ABSA selects one successful BSM arm to retain and measures the connected inner qubits in the X basis (shown), while all other inner qubits are measured in the Z basis.
        Once the measurement outcomes are communicated to Alice and Bob, a Bell pair is effectively shared between them.
    } \label{fig:rgs-overview}
\end{figure*}

In this work, we show that the recently proposed RGS architecture based on a building block called the half-RGS~\cite{naphan-rgs-protocol} enables the integration of conventional two-way purification with only a small overhead in generation rate (scaling with the half-RGS creation time) and a modest increase in resource requirements (additional quantum emitters equal to the number of purification rounds) at repeater stations.
Since the RGS’s loss-tolerant properties can be tuned by increasing the number of photons, the distribution process can be made near-deterministic, making optimistic entanglement purification protocols~\cite{hartmann-optimistic-purification, razavi-optimistic-purification-pumping, mobayenjarihani-optimistic-purification-qce} suitable with the RGS scheme.
In this approach, purification is performed immediately on the generated half-RGS primitives without waiting for heralding outcomes from measurement stations.
The coincidence measurements (denoting the success of purification) are then sent along with the heralding signals from link-level generation and Pauli frame corrections to be processed at the end nodes.
This enables purification between non-neighboring nodes and supports flexible purification scheduling with minimal overhead.
As a result, this allows the RGS scheme to benefit from scheduling strategies previously explored in memory-based quantum repeaters~\cite{krastanov-purification, poramet-quanta-extended, poramet-quanta-qcnc, haldar-purify-swap-schedule} and makes techniques that assume ideal memory conditions\cite{gidney-tetrationally-compact-purification, pattison-constant-rate-purification, ataides-constant-rate-purification-high-rate-codes, shi-stabilizer-purification} more relevant to practical all-photonic implementations.

The rest of this paper is organized as follows.
We first review the RGS-based all-photonic repeater architecture and the half-RGS primitive (Sec.~\ref{sec:rgs-overview}).
We then discuss the traditional two-way entanglement purification and the optimistic approach (Sec.~\ref{sec:purification}).
Later, we present our purification-enhanced framework, detailing how purification scheduling can be integrated into the RGS scheme and analyzing its impact on generation rates and resource requirements (Sec.~\ref{sec:rgs-with-purification}).
To demonstrate the flexibility of our approach, we begin by outlining the error modeling and assumptions used in our comparison setting (Sec.\ref{sec:error-modeling}), followed by numerical results based on a custom purification schedule that combines both link-level and non-neighbor purification; we compare these results against a baseline scheme using heralded end-to-end purification via entanglement pumping\cite{dur-repeater} (Sec.~\ref{sec:numerical-results}).
Finally, we conclude with open challenges and potential directions for future research (Sec.~\ref{sec:discussion}).

\section{Background: Overview of RGS Scheme}
\label{sec:rgs-overview}

The repeater graph state (RGS) consists of $2m$ inner qubits (orange qubits in~\ref{fig:rgs-overview}) and $2m$ outer qubits (pink qubits in~\ref{fig:rgs-overview}).
The inner qubits form a complete bipartite graph (biclique)\footnote{The inner qubits form a complete graph instead of a biclique in the original proposal of Azuma \emph{et al.}~\cite{azuma-rgs}}, where each partition contains $m$ qubits.
Each outer qubit is connected to one inner qubit, forming the arms of the RGS.
In a quantum network based on the RGS scheme, there are two types of intermediate nodes: RGS source stations (RGSS) and advanced Bell state analyzer stations (ABSA).
To distribute an entangled Bell pair between two end nodes, Alice and Bob, a path is established through a sequence of RGSSs and ABSAs, as illustrated in~\ref{fig:rgs-overview}.
At each RGSS, an RGS is generated and split into two halves along the bipartite partition.
These halves are sent to the adjacent ABSAs on either side.
Meanwhile, the end nodes (Alice and Bob) prepare and transmit half-RGS states entangled with their memories~\cite{naphan-rgs-protocol} to their respective neighboring ABSAs.
Each ABSA performs Bell state measurements (BSMs) on incoming pairs of outer qubits--one from each adjacent RGSS or end node.
A successful BSM establishes a link-level entangled pair between the neighboring RGSSs.
Once link-level entanglement is established across the entire path, single-qubit measurements are performed on the inner qubits to complete the entanglement swapping process, ultimately creating an end-to-end Bell pair between Alice and Bob.

The half-RGS~\cite{naphan-rgs-protocol} is a building block in which a group of $m$ inner qubits (one partition that will be sent to the same ABSA) is connected to an emitter qubit, known as the anchor qubit, and each inner qubit is also connected to an outer qubit (shown in~\ref{fig:rgs-overview}).
It can be deterministically generated using a small number of quantum emitters~\cite{buterakos-graph-generation, hilaire-rgs-optimizing-gen-time, li-entangled-photon-factory}.
This building block enables seamless integration with memory-based repeaters, by swapping the states of anchor qubits with a memory qubit, the memory qubit would hold one end of the established Bell pair.
If the end nodes are equipped with the same emitters capable of generating half-RGSs, the number of quantum memories required at the end nodes can be significantly reduced--especially in applications that demand deterministic Bell pair generation.
As the name suggests, two half-RGSs can be combined into a full RGS through an application of a controlled-phase (CZ) gate between their anchor qubits followed by X basis measurements on the anchor qubits.

In this paper, we make a clear distinction between quantum memories and quantum emitters based on their coherence time.
Quantum memories are assumed to have sufficiently long lifetimes to preserve qubits while classical messages are exchanged between distant end nodes.
In contrast, quantum emitters are short-lived qubits that only need to survive for durations comparable to the time required to generate a half-RGS.

We will show how leveraging the half-RGS structure enables the integration of entanglement purification into the RGS scheme.
Although the anchor qubits are implemented using short-lived emitter qubits, their role at RGSS nodes is functionally analogous to that of quantum memories in memory-based repeaters.
In addition to facilitating the generation of half-RGSs, the process of joining two half-RGSs into an RGS effectively realizes the entanglement swapping operation--if we interpret the graph state transformations in the order of outer qubits, followed by inner qubits, and finally the anchor qubits--even though the physical joining may occur before all qubits arrive at the ABSA.
This functional equivalence allows us to adapt purification techniques developed for memory-based architectures as we will show in later sections.

\section{Entanglement Purification}
\label{sec:purification}

\begin{figure*}[htp]
    \centering
    \includegraphics[width=\textwidth]{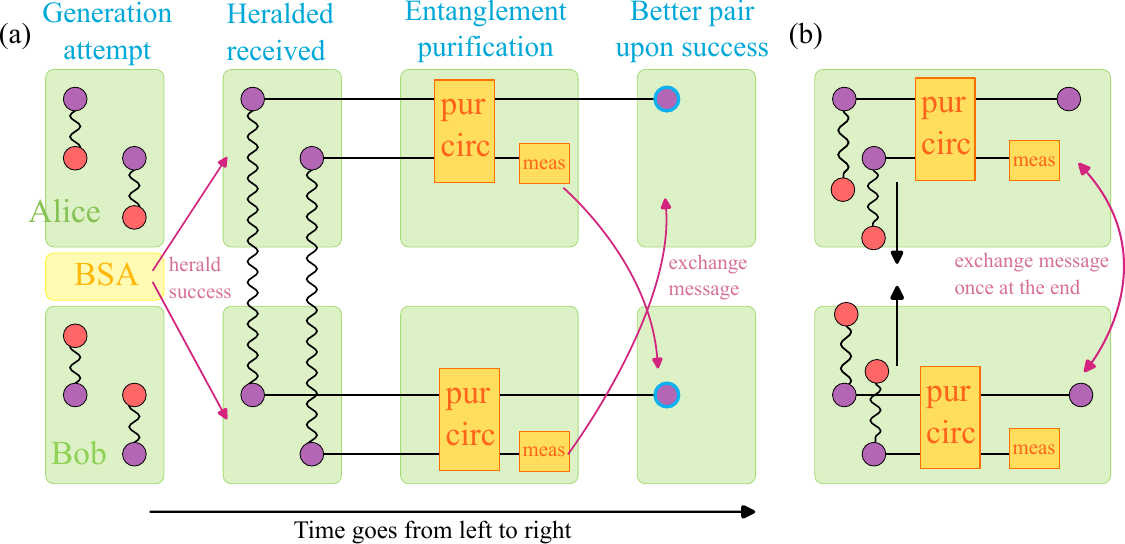}
    \caption{
    (a) Schematic of two-way heralded entanglement purification.
    The process begins with the probabilistic distribution of Bell pairs, followed by heralding signals from intermediate nodes confirming successful pair creation.
    A purification circuit is then applied, during which some Bell pairs are measured and consumed.
    The measurement results are exchanged between the two parties; if the outcomes agree (i.e., correlate or anti-correlate as expected), the remaining Bell pair is kept, resulting in improved fidelity.
    (b) Schematic of optimistic purification.
    In this approach, distribution and purification operations are performed immediately in sequence, without waiting for heralding confirmation.
    Classical outcomes are compared only once at the end, combining both the heralding of Bell pair generation and the results of purification in a single round of communication.}
    \label{fig:purification-schematic}
\end{figure*}

Entanglement purification is an essential technique for managing errors in quantum communication.
It raises the fidelity of generated Bell pairs to meet the threshold required by network applications running at end nodes.
In this section, we review the fundamental principles of entanglement purification and show how the optimistic variant can be integrated into the all-photonic repeater architecture we propose.

\subsection{Basics of entanglement purification}
\label{sec:purification:review}

In memory-based quantum repeaters, particularly in the first-generation quantum repeaters~\cite{muralidharan-repeater-generation}, entanglement purification is used to convert multiple noisy Bell pairs into a smaller number of higher-fidelity pairs.
This process serves as an error detection mechanism, where some pairs are sacrificed to detect the presence of errors.

Unlike error correction, which aims to detect and then fix errors to recover the original quantum state, error detection verifies whether an error has occurred.
If an error is detected, the faulty state is discarded and the entanglement generation process is restarted.
This approach is generally less resource-intensive and better suited for scenarios where the target state is known and relatively inexpensive to prepare.
Bell pairs, being symmetric and standard maximally entangled states, typically benefit more from error detection than correction.

There are two main strategies for implementing purification.
One is joint stabilizer measurement, where the same stabilizer operator is measured across multiple copies of Bell pairs~\cite{bennett-purification, deutsch-purification}.
Another involves encoding several Bell pairs into a logical state using a quantum error-correcting code~\cite{pattison-constant-rate-purification, ataides-constant-rate-purification-high-rate-codes, shi-stabilizer-purification}.
The latter method detects and potentially corrects errors by analyzing the error syndrome.
While this approach includes correction, it still functions similarly to error detection in practice, since we retain smaller numbers of logical Bell pair states than we started with.

In the quantum network setting, joint stabilizer measurement-based entanglement purification requires classical communication between the two parties to exchange measurement results.
This exchange ensures that both sides agree on whether purification was successful and whether to proceed with the retained pairs.
This communication-dependent process is known as \emph{two-way heralded purification}.
In contrast, certain error correction-based approaches can enable \emph{one-way purification}.
Since Bell pairs are symmetric, it is sufficient for only one party to receive the syndrome and apply local corrections.

In this paper, we focus on the two-way joint stabilizer measurement approach.
This method is conceptually straightforward and directly adaptable to our RGS scheme framework.
A schematic diagram illustrating 2-to-1 purification using joint stabilizer measurements is provided in Fig.~\ref{fig:purification-schematic}.
We will examine the propagation of errors and introduce the purification circuits used in this work in Sec.~\ref{sec:rgs-with-purification}.

\subsection{Optimistic purification}

Optimistic entanglement purification~\cite{hartmann-optimistic-purification, razavi-optimistic-purification-pumping, mobayenjarihani-optimistic-purification-qce}, also known as blind purification, is a variant of the two-way heralded purification approach. (A schematic of the approach is shown in ~\ref{fig:purification-schematic}).
In this method, both parties proceed with purification immediately, without waiting for confirmation that Bell pairs have been successfully established.
When multiple purification rounds are required, they are executed in sequence without intermediate heralding.
Only a single round of classical communication is exchanged at the end to report the outcome of both entanglement generation and purification.

While this approach may lower the overall success probability, and consequently the rate, of obtaining high-fidelity Bell pairs--especially when Bell pair generation is probabilistic--it provides significant advantages in high-latency settings.
Long-distance communication introduces delays, during which quantum memories can decohere.
As demonstrated in~\cite{razavi-optimistic-purification-pumping, mobayenjarihani-optimistic-purification-qce}, the cumulative effect of memory decoherence during multiple rounds of classical messaging can outweigh the benefits of traditional purification.
The optimistic approach mitigates this issue by minimizing memory idle time, leading to higher output fidelity even with fewer successful purification attempts.

In scenarios where the Bell pair generation rate is high or near-deterministic, optimistic purification is also especially well-suited.
Since Bell pairs are reliably created, there's little benefit in waiting for heralding signals before initiating purification.
This further reduces latency and maximizes throughput.
In our work, we adopt this strategy, as one key advantage of the RGS framework is the ability to generate Bell pairs in a near-deterministic manner.
In the next section, we demonstrate how optimistic purification can be seamlessly integrated into the all-photonic RGS framework by leveraging the architectural advantages provided by half-RGS building blocks.

\section{Purification-enhanced RGS Scheme}
\label{sec:rgs-with-purification}

With the necessary background established, we now show how the half-RGS building block enables entanglement purification to be naturally integrated into the RGS scheme.

\subsection{Modification to the RGS generation}
\label{sec:rgs-with-purification:generation}

\begin{figure*}
    \centering
    \includegraphics[width=\textwidth]{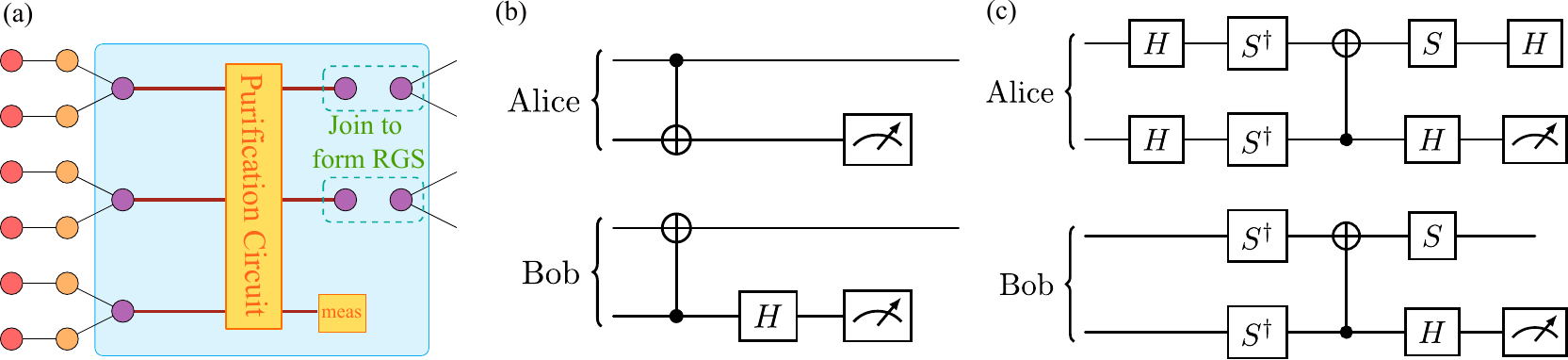}
    \caption{
        (a) Integration of entanglement purification into the RGS generation process.
        Instead of joining left and right half-RGSs immediately after their creation as in the original protocol~\cite{naphan-rgs-protocol}, multiple half-RGSs are generated on each side.
        Their anchor emitters serve as inputs to a purification circuit.
        After purification, the retained high-fidelity half-RGSs from the left and right are paired and joined to form complete RGSs.
        (Note that dark red lines represent the circuits or their evolution in time and do not represent entangled properties of graph states.)
        (b) Graph state purification circuits that measures the parity of stabilizer $Z_a X_b$ where $a$ represents the Alice's side and $b$ for Bob~\cite{dur-graph-state-purification, kruszynska-graph-state-purification} Top (bottom) qubits of Alice and Bob are Bell pairs in the two-qubit graph state.
        The top qubits for both side are retained after the purification circuit.
        The purification circuit for $X_a Z_b$ are similar, by exchanging the circuits that Alice and Bob locally performed.
        (c) Graph state purification circuit that measures the parity of $Y_a Y_b$ stabilizers.
    }
    \label{fig:purification-enhanced-rgs}
\end{figure*}

In prior work~\cite{naphan-rgs-protocol}, once half-RGSs are generated on both the left and right sides of a repeater, they are immediately joined to form a full RGS.
In our purification-enhanced scheme, we modify this process by delaying the joining step.
Instead of creating a single half-RGS per side, we generate multiple half-RGSs on each side.
This requires additional anchor emitters equal to the number of Bell pairs involved in the purification protocol.
Once all the half-RGSs are created, the anchor qubits--emitter qubits entangled with their respective inner qubits--are fed as inputs to the purification circuit.
After purification, the retained half-RGSs from both sides are paired and joined to form full RGSs.
This modified RGS generation process, incorporating purification at the half-RGS level, is illustrated in Fig.~\ref{fig:purification-enhanced-rgs}.

\subsection{Overhead Analysis of Purification-Enhanced RGS Scheme}
\label{sec:rgs-with-purification:overhead-analysis}

We now analyze the time overhead associated with our purification-enhanced RGS framework, with a focus on the memory coherence time required at the end nodes and the RGS source (RGSS) nodes.
In particular, we evaluate how long the end-node quantum memories and anchor emitter qubits at RGSS nodes must remain coherent to support network applications requiring high-fidelity Bell pairs.

We consider the following three scenarios for comparison: (1) Raw RGS scheme, where no purification is performed, as in the original protocol outlined in~\cite{naphan-rgs-protocol}; (2) Baseline purification, where entanglement purification is applied only at the end nodes after Bell pairs have been distributed; and (3) Optimistic purification, our proposed purification-enhanced RGS framework that integrates purification directly into the RGS generation process.
For simplicity, we assume that all scenarios employ the same tree encoding, allowing for direct comparison of timing overheads.
Let $\thrgs$ denote the time required to generate a single half-RGS.

\subsubsection{Raw RGS Scheme (no purification)}
In the original protocol, once half-RGSs are generated on both sides, they are immediately joined to form a full RGS.
The associated generation and memory times are given (respectively) as
\begin{align}
    \tau^{\text{(raw)}}_{\text{RGS}} &= \thrgs{} + \tau_{\text{join}}, \\
    t^{\text{(raw)}}_{\text{mem}} &\approx \thrgs{} + L_{\text{total}} / c,
\end{align}
where $\tau_{\text{join}}$ is the time to join two half-RGSs, $L_{\text{total}}$ is the total distance between end nodes, and $c$ is the speed of light in the communication channel.
Note that $\tau_{\text{join}}$ does not appear in the memory time, as it is negligible compared to the end-to-end transmission time, especially for large $L_{\text{total}}$. The memory coherence time must span the duration of half-RGS generation and the arrival of Pauli frame corrections from the remote end node.

\subsubsection{Baseline Purification at End Nodes}
In this approach, multiple Bell pairs are first distributed, and purification is performed afterward at the end nodes.
The memory coherence time must cover both generation and purification stages, and is given as
\begin{align}
    t^{\text{(p-base)}}_{\text{mem}} &\approx n_{\text{pur}} \thrgs{} + \tau_{\text{pur circ}} + 2L_{\text{total}} / c, \\
    t^{\text{(p-base)}}_{\text{mem}} &\approx n_{\text{pur}} \thrgs{} + 2L_{\text{total}} / c,
\end{align}
where $n_{\text{pur}}$ is the number of Bell pairs used per purification attempt, and $\tau_{\text{pur circ}}$ is the time to apply the purification circuit.
The generation time $\tau_{\text{RGS}}$ remains the same as in the raw scheme.
However, the overall memory requirement is higher due to the need to wait for heralded confirmation before applying purification, which doubles the communication latency.
While we assume purification circuit time is negligible compared to transmission, multiple purification rounds would further increase the latency.

\subsubsection{Proposed Optimistic Purification Scheme}
Our enhanced approach integrates purification directly into the RGS generation process using optimistic purification.
The RGS generation and memory coherence times are given (respectively)  as 
\begin{align}
    \tau^{\text{(p-opt)}}_{\text{RGS}} &= n_{\text{pur}}\thrgs{} + \max \{\tau_{\text{pur circ}} + \tau_{\text{join}},  n_{\text{pur}}\tau_{\text{join}}\}, \\
    t^{\text{(p-opt)}}_{\text{mem}} &\approx n_{\text{pur}}\thrgs{} + \tau_{\text{pur circ}} + L_{\text{total}} / c.
\end{align}
We take the maximum in $\tau^{\text{(p-opt)}}_{\text{RGS}}$ because some RGSS nodes may need to prepare multiple raw RGSs in sequence (if not selected to perform half-RGS purification), while others perform purification on several half-RGSs before joining the retained ones.
Despite this added complexity, the memory time scales significantly better than in the baseline case, particularly in high-latency networks.
If the half-RGS generation time $\thrgs$ is small relative to the classical message transmission time $L_{\text{total}}/c$, the memory requirements of our enhanced scheme become comparable to those of the unpurified (raw) RGS protocol.
While our approach reduces memory idling time, it does so at the cost of a reduced distribution rate: each purification attempt consumes $n_{\text{pur}}$ half-RGSs, decreasing the trial rate by a factor of $n_{\text{pur}}$.
However, compared to the raw approach, our scheme enables end nodes to reach the required fidelity threshold more quickly, even with fewer quantum memories, assuming a suitable purification protocol is employed.
Importantly, the anchor qubits at RGSS nodes only need to remain coherent for the duration of the RGS generation process--shifting the primary memory burden to the end nodes.
Since the fidelity of the generated Bell pairs is often more critical than the raw distribution rate, overall performance is governed by the rate of high-fidelity Bell pair generation.
In the next section, we show that our purification-enhanced RGS scheme significantly improves this performance compared to the baseline, while the raw scheme fails to meet the fidelity threshold required by practical applications.

\section{Analytical Model}
\label{sec:error-modeling}

In this section, we demonstrate the flexibility and evaluate the performance of the purification-enhanced RGS scheme through numerical simulations.
We show the flexibility by applying the purification on the half-RGS both at the link-level and at some non-neighbor nodes.
For the scope of the current paper, we only focus on the fidelity of the distributed Bell pairs and the rate of given minimum fidelity.
We do not fully investigate the optimization of purification protocols selection and scheduling. 

\subsection{Scope and Assumptions}

We consider a repeater chain of 10 hops, with 9 intermediate RGS source nodes (RGSS) and two end nodes equipped with quantum memories (Alice and Bob). We do not detail the generation process of half-RGS, and instead directly apply the effective errors (from inner qubits) to the anchor qubits, as modeled in~\cite{naphan-rgs-protocol-tqe}. The half-RGS are generated with parameters tuned to achieve near-deterministic success rates ($\geq 0.999$).
Relevant simulation assumptions and parameters are as follows:
\begin{itemize} 
\item Nodes are spaced 2 kilometers apart, and the loss in the optical transmission channel is 0.2dB per kilometer (i.e., each photon needs to travel 2km from source to measurement nodes). This choice is arbitrary and is chosen for modeling convenience.
\item No other sources of photon loss in the system (e.g., no couple loss, no failure of photon emissions, etc.). 
\item Each half-RGS is constructed using 18 arms and a tree-encoding with branching parameters $(16, 14, 1)$, ensuring a high end-to-end generation probability ($ \geq 0.999$) at this given loss rate. 
\item Photons are affected by depolarizing noise, which is applied uniformly across all photons in the system. 
\item Quantum gates and measurements are assumed noiseless. 
\item Memories are assumed ideal (i.e., no decoherence). 
\end{itemize}
We justify the ideal assumptions for quantum operations and memories as follows.
In emitter-based graph state generation approaches~\cite{buterakos-graph-generation, li-entangled-photon-factory}, the gate and measurement errors can be effectively modeled as additional noise on the emitted photons, simplifying the simulation without loss of generality.
Regarding memories, recent advances in quantum error correction have enabled logical qubits with significantly extended coherence times~\cite{google-surface-code-one-qubit, google-surface-code-beyond-threshold, quantinuum-microsoft-logical-below-threshold, bluvstein-quera-qec}.
It is reasonable to assume that practical implementations at the end nodes will incorporate such technologies.
This assumption actually gives an advantage to the baseline setting, where purification is applied on stored logical entangled pairs.
Performing purification on encoded logical memories typically involves additional operations, which can introduce more error or delay.
In contrast, Bell pairs in our purification-enhanced framework can be encoded from a single pair after purification is performed on the unencoded Bell pairs, thus only single-qubit Pauli operations are needed for Pauli frame correction.
Single-qubit Pauli gates are cheap or even “free” in many error-correcting codes~\cite{gottessman-qec-book}.

\subsection{Purification Circuits and Scheduling}

We restrict our noise model to Pauli errors, which is both practical and justified in the context of logical qubits (for our ideal memory assumption).
Most error-correcting codes transform general noise into effective Pauli channels through syndrome measurements, making this a natural abstraction~\cite{gottessman-qec-book}.
Both the baseline and purification-enhanced settings use the same purification strategy: three variants of the 2-to-1 stabilizer-based purification protocol.
These operate by measuring joint stabilizers on two copies of two-qubit graph states.
The three circuits (shown in Fig.~\ref{fig:purification-enhanced-rgs}) measure the stabilizers $Z_a X_b$, $X_a Z_b$, and $Y_a Y_b$, respectively.
The first two purification circuits were proposed in~\cite{dur-graph-state-purification, kruszynska-graph-state-purification} while we construct the $YY$ variant ourselves.

Because two-qubit graph states (Bell pairs) are symmetric, Pauli error states can be interpreted in two equivalent ways: one where all errors affect just one qubit (e.g., qubit $b$), and another where all errors are $Z$-type and can affect either or both qubits.
We adopt the latter interpretation, as it simplifies error propagation analysis and clarifies the purification circuit behavior as it was used by protocols in~\cite{naphan-rgs-protocol, naphan-rgs-protocol-tqe} for calculating Pauli frame corrections.
The three representative Pauli error cases on the two-qubit graph state can be mapped as: 
\begin{enumerate}
\item $I_a X_b = Z_a I_b$
\item $I_a Z_b = I_a Z_b$
\item $I_a Y_b = Z_a Z_b$
\end{enumerate}
Each purification circuit detects different types of $Z$ errors:
\begin{itemize} 
\item The $ZX$ stabilizer detects $Z$ errors on qubit $b$, and fails if exactly one input pair is affected by $IZ$ or $ZZ$ errors.
However if both pairs are affected by $IZ$ or $ZZ$ errors, the protocol would yield a false positive.
\item The $XZ$ stabilizer behaves analogously, but detects $Z$ errors on qubit $a$.
\item The $YY$ stabilizer detects an odd number of $Z$ errors (one or three across the four qubits involved), and only gives a false positive if an even number of $Z$ errors are present.
\end{itemize}

\subsection{Error Modeling and Effective Noise in RGS}

Our purification schedule for the baseline setting is based on the entanglement pumping protocol from~\cite{gidney-tetrationally-compact-purification}.
We perform repeated rounds of purification in the sequence: $YY$, $ZX$, $YY$, $XZ$, and continue until the fidelity exceeds a chosen threshold.
The intuition behind this sequence is as follows.
Although photons are subjected to symmetric depolarizing noise--implying equal probabilities for $IZ$, $ZI$, and $ZZ$ errors--the error contributions to the anchor qubits, originating from the inner qubits of the half-RGS, exhibit a bias in their distribution.
In particular, these inherited errors produce independent $Z$ errors on each end node, making $IZ$ and $ZI$ more prevalent than $ZZ$.
As a result, the $YY$ purification circuit, which detects an odd number of $Z$ errors across the qubits involved, is more likely to identify and eliminate such errors in the early rounds, thereby accelerating fidelity improvement when used at the beginning of the schedule.

Given our Pauli noise model, we can represent the mixed state of the Bell pairs over four possible Bell pair states as a vector of length 4 given by
\begin{align}
    [w, x, y, z]^T \coloneq &w\ket{\psi_{II}}\bra{\psi_{II}} + x\ket{\psi_{ZI}}\bra{\psi_{ZI}} \nonumber\\
        &+ y\ket{\psi_{ZZ}}\bra{\psi_{ZZ}} + z\ket{\psi_{IZ}}\bra{\psi_{IZ}},
\end{align}
where $\ket{\psi_{P}}$ denotes the two-qubit graph state affected by Pauli operator $P$, and the vector sums to 1 ensuring a valid density operator.
We denote this vector by a shorthand notation $\mathbf{e} = [w, x, y, z]^T$.

The probability of success for each purification circuit is given (as a function of two error vectors) by
\begin{align}
    P_{ZX}\left(\mathbf{e_1}, \mathbf{e_2}\right) &= (w_1 + x_1) (w_2 + x_2) + (z_1 + y_1) (z_2 + y_2), \\
    P_{XZ}\left(\mathbf{e_1}, \mathbf{e_2}\right) &= (w_1 + z_1) (w_2 + z_2) + (x_1 + y_1) (x_2 + y_2), \\
    P_{YY}\left(\mathbf{e_1}, \mathbf{e_2}\right) &= (w_1 + y_1) (w_2 + y_2) + (x_1 + z_1) (x_2 + z_2).
\end{align}
Conditioned on success, the purified error vectors are
\begin{align}
    \text{Pur}_{ZX}\left(
        \begin{bmatrix}
           w_1 \\
           x_1 \\
           y_1 \\
           z_1
        \end{bmatrix},
        \begin{bmatrix}
           w_2 \\
           x_2 \\
           y_2 \\
           z_2
        \end{bmatrix}
    \right) &= \frac{1}{P_{ZX}(\mathbf{e_1}, \mathbf{e_2})}\begin{bmatrix}
           w_1  w_2 + z_1  z_2 \\
           z_1  w_2 + w_1  z_2 \\
           x_1  y_2 + y_1  x_2 \\
           x_1  x_2 + y_1  y_2 
        \end{bmatrix}, \\
    \text{Pur}_{XZ}\left(
        \begin{bmatrix}
           w_1 \\
           x_1 \\
           y_1 \\
           z_1
        \end{bmatrix},
        \begin{bmatrix}
           w_2 \\
           x_2 \\
           y_2 \\
           z_2
        \end{bmatrix}
    \right) &= \frac{1}{P_{XZ}(\mathbf{e_1}, \mathbf{e_2})}\begin{bmatrix}
           w_1  w_2 + x_1  x_2 \\
           z_1  z_2 + y_1  y_2 \\
           z_1  y_2 + y_1  z_2 \\
           x_1  w_2 + w_1  x_2 
        \end{bmatrix}, \\
    \text{Pur}_{YY}\left(
        \begin{bmatrix}
           w_1 \\
           x_1 \\
           y_1 \\
           z_1
        \end{bmatrix},
        \begin{bmatrix}
           w_2 \\
           x_2 \\
           y_2 \\
           z_2
        \end{bmatrix}
    \right) &= \frac{1}{P_{YY}(\mathbf{e_1}, \mathbf{e_2})}\begin{bmatrix}
           w_1  w_2 + y_1  y_2 \\
           x_1  z_2 + z_1  x_2 \\
           y_1  w_2 + w_1  y_2 \\
           x_1  x_2 + z_1  z_2 
        \end{bmatrix}.
\end{align}

To account for the Bell state measurement (BSM) process applied to the outer qubits, the error vector of the post-BSM state is computed as
\begin{align}
    \text{BSM}_{ZX}\left(
        \begin{bmatrix}
           w_1 \\
           x_1 \\
           y_1 \\
           z_1
        \end{bmatrix},
        \begin{bmatrix}
           w_2 \\
           x_2 \\
           y_2 \\
           z_2
        \end{bmatrix}
    \right) &= \begin{bmatrix}
           w_1 w_2 + x_1 x_2 + y_1 y_2 + z_1 z_2 \\ 
           w_1 x_2 + x_1 w_2 + z_1 y_2 + y_1 z_2 \\ 
           w_1 y_2 + y_1 w_2 + x_1 z_2 + z_1 x_2 \\
           w_1 z_2 + z_1 w_2 + x_1 y_2 + y_1 x_2 
        \end{bmatrix}.
\end{align}
Note that the BSM employed in this work is a variant of rotated BSM typically used in RGS scheme literature, i.e., a CZ gate on the two input qubits followed by single qubit X-basis measurements on them.

\subsection{Evaluation Methodology}

\begin{figure*}
    \centering
    \includegraphics[width=\textwidth]{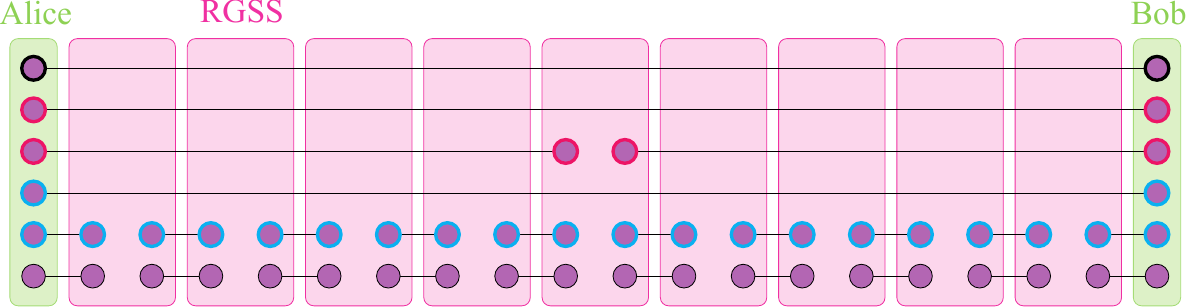}
    \caption{
        Illustration of flexible purification scheduling enabled by our purification-enhanced RGS framework.
        Using five half-RGSs per side at each hop (bottom line represents a copy), three end-to-end Bell pairs are created via different purification schedules, resulting in varying fidelities.
        These pairs then undergo further purification at the end nodes to yield a single high-fidelity Bell pair.
        The first is constructed by purifying ($YY$) each link-level connection and stitching the purified links across the full path (blue border).
        The second is formed by stitching link-level pairs into two copies of two 5-hop segments, purifying ($YY$) within each segment, and then connecting them (pink border).
        The third is created by directly stitching raw link-level pairs without intermediate purification (black thick border).
        These three pairs are then used in a pumping-like purification sequence: the first and second are purified (with the second as the sacrificial pair with $ZX$), followed by purification between the result and the third (with the third as the sacrificial pair with $YY$).
        The final output is a high-fidelity end-to-end Bell pair, significantly improved over the unpurified case upon success.
        }
    \label{fig:flexible-scheduling}
\end{figure*}

To track the effective error vectors in the RGS scheme, we condition on the successful generation of all one-hop Bell pairs between the anchor qubits--that is, we only simulate events in which entanglement distribution succeeds.
As shown in Fig.~\ref{fig:rgs-overview}, once the outer qubit arms leading to successful BSMs are selected and the measurement bases of the inner qubits are chosen, the sequence of measurements, error propagation, and Pauli frame corrections can be freely rearranged without affecting the outcome.
This allows us to model each half-RGS as an effective two-qubit graph state between an anchor and an outer qubit, with effective noise from the inner qubits directly applied to the anchors.
To simulate this noise, we use the average logical error probability--an approach also adopted in~\cite{azuma-rgs, hilaire-rgs-optimizing-gen-time}--which provides a good approximation, especially with many inner qubits and long repeater chains.
Since our goal is to understand fidelity trends rather than fault-tolerance thresholds, this approximation is sufficient.

The simulation proceeds as follows: we begin with a two-qubit graph state between each anchor and its corresponding outer qubit.
Depolarizing noise is applied to the outer qubits, followed by Bell state measurements (BSMs) and appropriate Pauli frame corrections on the anchors.
Effective error channels from logical measurements on the inner qubits are then incorporated by updating each anchor’s error vector according to the average logical error probabilities.
The average logical error probabilities for $X$ and $Z$ basis measurements are taken from~\cite{hilaire-rgs-optimizing-gen-time} and applied according to the error propagation rules in~\cite{naphan-rgs-protocol-tqe}.
The inner qubits induce independent $Z$-type error channels on each anchor: with probability given by the average logical error, we update the error vector as if up to $(m-1)$ $Z$ errors occurred from logical $Z$ measurements, and one additional $Z$ from the logical $X$ measurement.

While real systems would apply Pauli corrections only at the end nodes~\cite{naphan-rgs-tutorial}, applying local corrections at intermediate steps yields equivalent results.
This approach also simplifies explanation and error tracking, especially when including purification circuits.

After these steps, we apply purification protocols depending on the scenario under consideration.
In the baseline purification approach, we perform Bell-state measurements (BSMs) between all anchor qubits that are not part of the end nodes to create end-to-end Bell pairs, and then apply purification circuits using these Bell pairs.

In contrast, for our proposed purification scheme integrated with RGS creation, we first generate multiple copies of each half-RGS, perform purification on them locally, and then join the purified copies into a full RGS at the appropriate points in the network.
Additional rounds of purification may then be performed between the anchors, depending on the chosen purification schedule.
We now describe the full set of calculations required to determine and compare end-to-end fidelities across different frameworks.

\section{Numerical Results}
\label{sec:numerical-results}

In the purification-enhanced scheme, purification is scheduled at multiple stages along the connection path rather than being deferred to the end nodes.
Specifically, we construct three end-to-end Bell pairs using a combination of local and mid-path purification: one pair is obtained by performing link-level purification with $YY$ stabilizer and stitching the purified links, another is purified at the midpoint of the path with $YY$ stabilizer and then extended outward, and the third is an unpurified raw Bell pair (as shown in Fig.~\ref{fig:flexible-scheduling}).
These three pairs are then combined and further purified at the end nodes to yield a final high-fidelity connection.
The exact sequence we performed is $ZX$ stabilizer purification is applied between the first and second pair (with the second being measured), followed by purification with $YY$ stabilizer with the third pair (with the third being measured).
This staged purification schedule highlights the flexibility of our approach in exploiting purification opportunities throughout the distribution process.

\begin{figure}[!htp]
    \centering
    \includegraphics[width=\columnwidth]{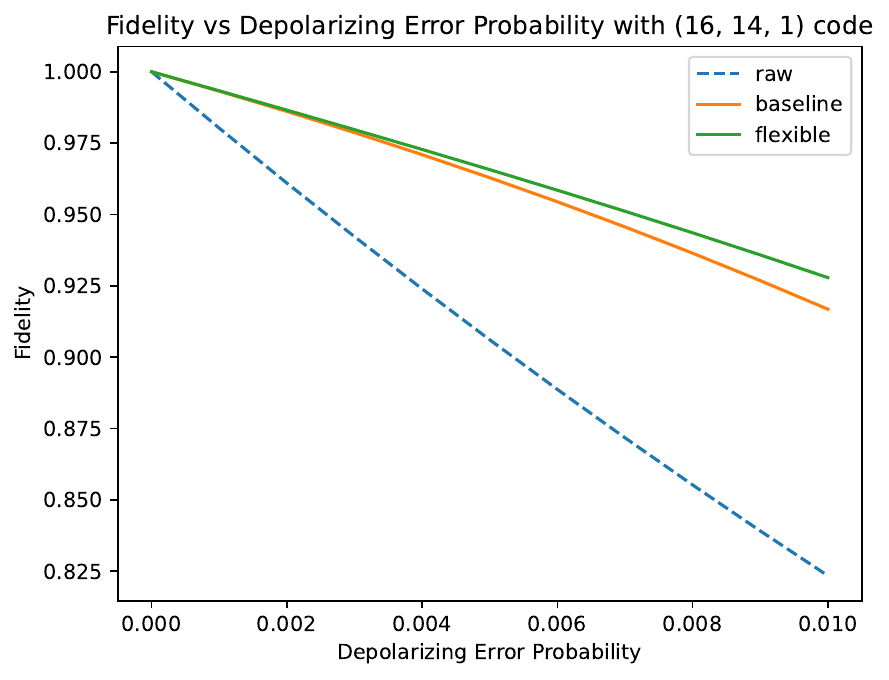}
    \caption{End-to-end fidelity comparison between unpurified Bell pairs, baseline purification using direct end-to-end Bell pairs, and our purification-enhanced scheme.
    Both of the purified schemes consume five half-RGSs per instead of one in the unpurified setting.}
    \label{fig:fidelity-plot}
\end{figure}
\begin{figure}[!htp]
    \centering
    \includegraphics[width=\columnwidth]{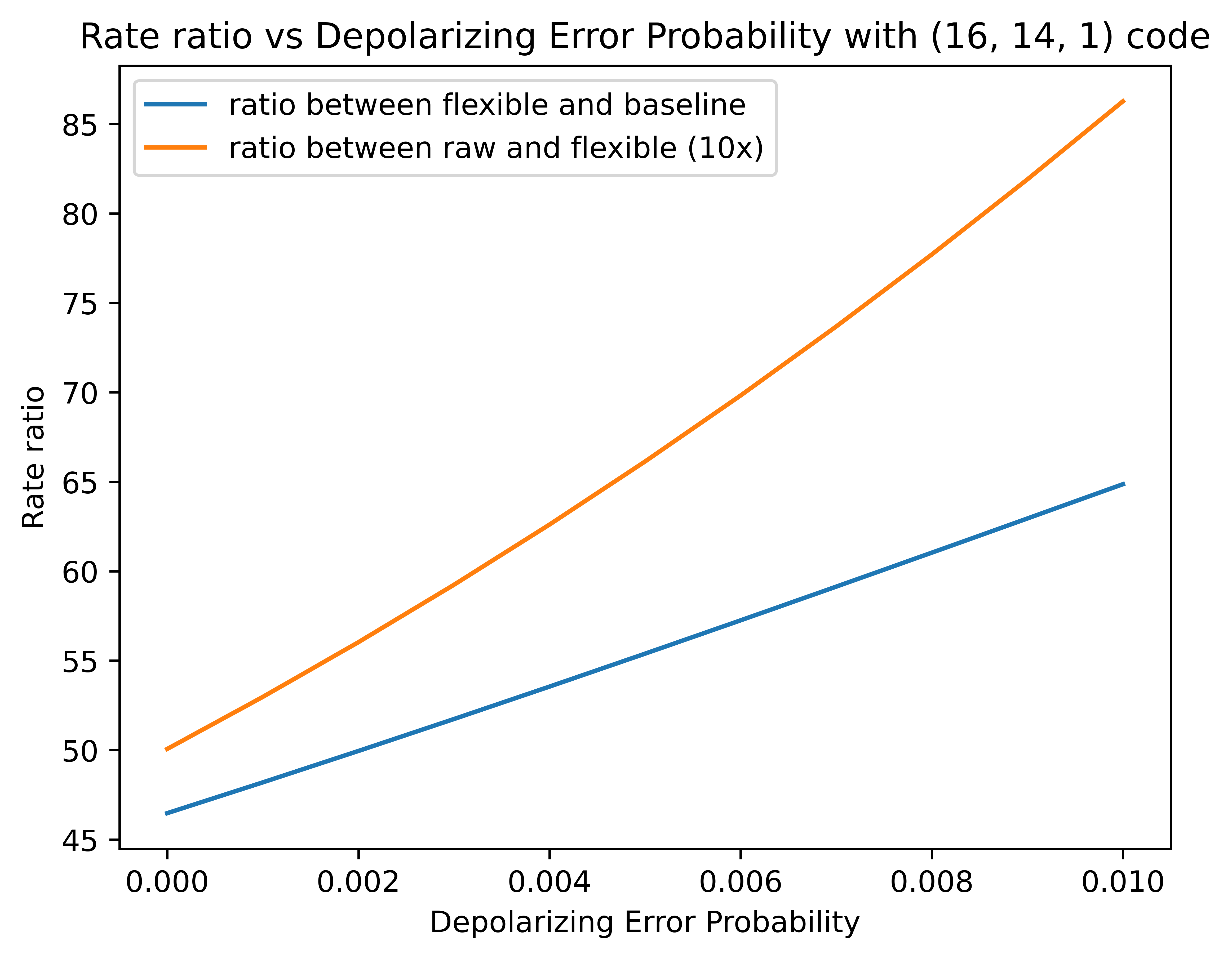}
    \caption{
        Relative distribution rates under different purification strategies. 
        The ratio between raw and purification-enhanced as well as between baseline and purification-enhanced is shown. The ratio between raw and purification-enhanced are scaled by a factor of 10 for easier comparison.
    }
    \label{fig:rate-ratio-plot}
\end{figure}

Fig.~\ref{fig:fidelity-plot} and Fig.~\ref{fig:rate-ratio-plot} together illustrate the trade-offs between fidelity and rate across the three purification strategies.
Both the baseline and the purification-enhanced schemes significantly improve the end-to-end fidelity compared to the unpurified case, achieving fidelities above 0.9 across the range of depolarizing noise parameters considered ($0 - 0.01$).
Notably, the purification-enhanced scheme consistently achieves higher fidelity than the baseline approach.
This improvement arises from its ability to suppress errors earlier in the repeater chain through intermediate purification steps, which limits the accumulation and propagation of noise before the final purification among end-to-end Bell pairs.

In terms of rate, the purification-enhanced scheme also demonstrates a clear advantage.
As shown in Fig.~\ref{fig:rate-ratio-plot}, it outperforms the baseline strategy by a factor ranging from approximately 45 to 65, depending on the noise level.
This substantial gain is primarily due to the avoidance of classical communication multiple rounds of end nodes communications over the full path, which are required in the baseline scheme for coordinating entanglement pumping across long distances.

In contrast, our framework pushes these exchanges of purification results to the very end by adopting the optimistic purification.
It is important to emphasize that both purified schemes in this comparison use the same number of half-RGSs per hop--five per side per RGSS and end nodes--ensuring that the observed improvements are not the result of differing resource consumption but rather of protocol design.
The purification-enhanced scheme exploits the structure of repeater graph states to embed purification within the entanglement generation process, enabling multiple stages of error suppression mechanism that enhance fidelity while simultaneously reducing time overhead.

These results are obtained without fully optimizing the purification schedule, suggesting that further improvements should be achievable through optimized scheduling strategies~\cite{krastanov-purification} tailored to specific network topologies or error profiles.
Overall, this highlights the potential of our purification-enhanced RGS framework to simultaneously meet the dual goals of high fidelity and high entanglement distribution rate, even in the presence of noise.

\section{Discussion}
\label{sec:discussion}

This work addresses a long-standing challenge in all-photonic quantum repeater design: how to naturally integrate entanglement purification into the all-photonic repeater graph state (RGS) architecture without compromising its defining advantages--fast, near-deterministic generation and memory-free operation at intermediate nodes.
We introduce a purification-enhanced RGS framework based on half-RGS building blocks, enabling flexible purification scheduling across both link-level and long-range segments of a connection path.

By incorporating entanglement purification into the RGS generation process, our approach unlocks the ability to implement other purification schedules that have been studied for memory-based repeaters into all-photonic repeaters naturally, such as the principle of ``purify-then-swap,'' which has been shown to perform well across many parameter regimes.

Our analysis here is meant to demonstrate feasibility rather than optimality.
We present a non-trivial yet manually selected schedule to showcase the flexibility of our framework.
Selecting optimal purification schedules or tuning RGS parameters for specific operational conditions remains an open problem and is a promising avenue for future research.

Several promising directions remain open.
For instance, constant-rate purification protocols~\cite{pattison-constant-rate-purification, ataides-constant-rate-purification-high-rate-codes, shi-stabilizer-purification} have been proposed in recent years, and it is worth investigating whether they can be adapted to our framework.
Another is how our method could leverage multiple successful Bell state measurements on unused RGS arms to further improve rates and efficiency, as explored in~\cite{bikun-li-generalized-rgs}.
Or utilizing purification outcome statistics, which could support real-time estimation and monitoring of error channels~\cite{joshua-distimator}.
This functionality is crucial for dynamic adaptation in practical quantum networks, enabling protocols to respond to evolving noise conditions.

Because our approach relies on multiple half-RGSs, it motivates continued investigation into more efficient generation techniques.
Recent advances in graph state creation~\cite{ghanbari-hoi-kwong-rgs-optimization} may offer pathways to reduce overheads and enhance fidelity.

Finally, studying our framework with network-level simulations~\cite{cocori-quisp, wu2021sequence, coopmans2021netsquid}--including traffic, routing, and control protocols--will be crucial for evaluating its performance in realistic deployments.
We hope this work encourages further research into purification scheduling, system integration, and scalable architectures for high-fidelity quantum communication.

\section*{Code Availability}

The numerical results presented in this paper and the analytical method are available at:
\url{https://github.com/Naphann/repeater-graph-state-protocol-based-on-half-RGS/tree/main}.

\bibliographystyle{IEEEtran.bst}
\bibliography{IEEEabrv, bibfile}

\end{document}